\documentclass[a4paper,11pt]{article}

\usepackage[english]{babel}
\usepackage[utf8x]{inputenc}
\usepackage{siunitx}

\usepackage{amsmath,amssymb}
\usepackage{graphicx}
\usepackage[colorlinks=true, allcolors=black]{hyperref}
\usepackage{geometry}
\usepackage{natbib}
\usepackage{graphics}
\usepackage{setspace}
\usepackage{ulem}

\newtheorem{remark}{Remark}

\begin{document}

\title{Workers' Incentives and the Optimal Taxation of AI\thanks{We would like to thank Pawe{\l} Struski (GRAPE, Warsaw, Poland) for his valuable comments and suggestions and for his input in the formulation of the first version of the model. We acknowledge the research funding from the National Science Centre Poland through the grant ``Will Artificial General Intelligence Bring Extinction or Cornucopia? Modeling the Economy at Technological Singularity'' (OPUS 26 No. 2023/51/B/HS4/00096).}}

\author{Jakub Growiec\thanks{SGH Warsaw School of Economics, Poland and CEPR Research and Policy Network on AI.}  \and Klaus Prettner\thanks{WU Vienna, Austria.} \and Maciej Szkr{\'o}bka\thanks{SGH Warsaw School of Economics, Poland.}}

\maketitle


\begin{abstract}
\noindent We characterize the optimal tax policy in an economy with human manual and cognitive labor, physical capital, and artificial intelligence (AI). Extending the dynamic taxation setup of \cite{Slavik2014}, we find that it is optimal to start taxing AI 
when cognitive workers start to consider switching to manual jobs. This threshold may be crossed once AI becomes sufficiently capable in substituting humans across cognitive tasks.
\vspace{0.2mm} 
\end{abstract}

\newpage

\section{Introduction}
\label{sec:intro}

The rapid development of artificial intelligence (AI) offers new opportunities for the automation of non-routine cognitive tasks \citep{AcemogluRestrepo2018a,GPTs,GmyrekEtal2025}. This new wave of automation builds on top of the previous wave of routine task automation \citep{acemoglu_autor_2011,Autor2013}. While empirical evidence shows that the former wave has increased wage inequality between routine and non-routine workers \citep[][]{Autor2003}, AI seems to have the opposite effect \citep[][]{Minniti2025}. In any case, however, both waves of automation contribute to the shift of factor income from workers towards capital owners, which has led to calls for a (higher) taxation of capital. 


In this note we analyze whether taxing AI is a reasonable policy. Extending the dynamic taxation setup of \cite{Slavik2014}, we identify the socially optimal taxation of traditional capital and AI. Our results show that taxing AI becomes optimal when cognitive workers start to consider switching to manual jobs. Given the rapid progress in AI capabilities and cognitive automation, this condition could soon be fulfilled. 

\section{The Model}
\label{sec:model}

We consider an economy populated by a continuum of agents of measure one. Following \cite{Slavik2014}, each agent either exhibits an advantage in cognitive or in manual tasks depending on agent type \(h \in H =\{c,m\}.\) The fraction of type-\(h\) agents is \(\pi_h\) and the type cannot be changed. Additionally, there are two types of capital: traditional capital (machines, assembly lines, structures, etc.) denoted by \(K\), and AI capital denoted by \(AI\). This four-factor setup resembles the AI and skill premium setup \citep{Bloometal2025AIFinal}, the equipment, structures, and skill premium setup \citep{McAdamWillman2018}, and the hardware--software framework \citep{GrowiecEtal2024}. 

\subsection{Production}
Each agent of type \(h\)  works \(l_h\) units of time and  produces \(l_h \cdot z_h\) units of effective labor, while the aggregate amount of effective \(h\)-type labor is given by \(L_h=\pi_h l_h z_h \). 
 The production function is \(Y=F(L_c,L_m,K,AI)\) and total wealth of the economy amounts to \(\tilde{F}=Y+(1-\delta_K)K+(1-\delta_{AI})AI\), where \(\delta_K\) and \(\delta_{AI}\) denote the depreciation rates. 
 Type-\(h\) workers earn the wage \(w_h\) given by
\begin{equation}
w_h=\frac{\partial F}{\partial L_h}\cdot z_h
\label{eq:1}.
\end{equation}
Following \cite{Slavik2014}, the complementarity between traditional capital $K$ and labor is higher for cognitive labor $L_c$ than for manual labor $L_m$. For $AI$ we assume the reverse: the degree of complementarity between $AI$ and labor is lower for cognitive labor $L_c$ than for manual labor $L_m$. Both assumptions are intuitive, supported by empirical evidence \citep[][]{Autor2003, Minniti2025} and well-grounded in theory \citep[][]{Bloometal2025AIFinal}. 
 Finally, we also apply the conventional assumption that the cognitive wage premium (the ratio of the cognitive wage to the manual wage) is strictly increasing with the amount of effective manual labor and strictly decreasing with the amount of effective cognitive labor. Summarizing the assumptions formally, we have 

\bigskip\noindent \textbf{Assumption 1}: 
\[
\frac{\partial F}{\partial L_c} \bigg{/} \frac{\partial F}{\partial L_m} \quad \text{strictly increases with } K.
\]

\noindent \textbf{Assumption 2}: 
\[
\frac{\partial F}{\partial L_c} \bigg{/}\frac{\partial F}{\partial L_m} \quad \text{strictly decreases with } AI.
\]

\noindent \textbf{Assumption 3}: 
\[
\frac{\partial F}{\partial L_c} \bigg{/}\frac{\partial F}{\partial L_m} \quad \text{strictly decreases with \(L_c\) and strictly increases with \(L_m\)}.
\]

Importantly, Assumptions 1-3 are so general that they do not determine whether any two factors are gross substitutes or gross complements. Instead, our assumptions allow both AI being a tool that is complementary to human cognitive work, and AI being an autonomous agent that can replace some or all human cognitive work.

\subsection{Preferences, Feasibility and the Social Planner Problem}

The preferences of type-\(h\) agents are characterized by
\[\sum_{t=0}^\infty \beta^t\bigg[u(c_{h,t})-\nu(l_{h,t})\bigg],\]
where \(\{c_{h,t},l_{h,t}\}_{t=0}^\infty\) is a sequence of consumption and labor supply, \(\beta\in(0,1)\) is the discount factor and \(u,\nu\colon\mathbb{R}_{+}\to \mathbb{R}\) are twice differentiable functions such that \(u',\nu',\nu''>0\) and \(u''<0\). 

Following \cite{Slavik2014}, we define an allocation as
\[\mathcal{X}=\{(c_{h,t},l_{h,t})_{h\in H},L_{m,t},L_{c,t},K_t,AI_t\}_{t=0}^\infty.\]
This allocation is feasible if and only if for every period \(t\geq0\) it satisfies:
\begin{equation}
    \sum_{h\in H}c_{h,t}+K_{t+1}+AI_{t+1}+G_t\leq \tilde{F}_t, \label{eq:2}
\end{equation}
where \(\{G_t\}_{t=0}^\infty\) is a sequence of government spending and \(\tilde{F}_t=\tilde{F}(L_{c,t},L_{m,t},K_t,AI_t)\). An allocation is incentive-compatible if it satisfies
\begin{equation}
\sum_{t=0}^{\infty}\beta^t\bigg[u(c_{h,t})-\nu(l_{h,t})\bigg]\geq \sum_{t=0}^\infty\beta^t\bigg[u(c_{j,t})-\nu\bigg(\frac{l_{j,t}w_{j,t}}{w_{h,t}}\bigg)\bigg], \label{eq:3}
\end{equation}
where \(j\) is the corresponding complement to \(h\) in \(H\). This incentive compatibility constraint (ICC) prevents agents of one type from mimicking the other type. It states that the discounted lifetime utility of type-$h$ agents when supplying the amount of labor designated to this type is weakly greater than the discounted lifetime utility of that agent when supplying the amount of labor that would be required to get the consumption level of the other type. 

The problem of the social planner who does not prefer any agent is
\[\max_{\mathcal{X}}\sum_{h\in H}\pi_h\sum_{t=0}^\infty \beta^t\bigg[u(c_{h,t})-\nu(l_{h,t})\bigg]\]
subject to (\ref{eq:1}), (\ref{eq:2}), (\ref{eq:3}), \(K_0\leq K_0^*\), \(AI_0\leq AI_0^*\), and \(L_h=\pi_h l_hz_h\). It leads to a constrained efficient solution that we denote by asterisks. 

\section{Optimal Taxation of Capital and AI}
\label{sec:tax}
\cite{Slavik2014} show that if the ICC binds for skilled agents, the optimal tax on equipment capital is strictly greater than the optimal tax on structure capital. In our framework, which contrasts cognitive workers with manual ones, where one group is not necessarily more skilled than the other, the results also crucially depend on the ICC. Below, we first use the original formulation (Assumption 4: ICC is binding for cognitive workers) and then its modified version (Assumption $4^\prime$: ICC is binding for manual workers), where the latter becomes relevant with sufficiently advanced AI. 

\subsection{ICC Binding for Cognitive Workers}

\noindent \textbf{Assumption 4:} 
 The ICC binds for cognitive workers $h=c$ and is slack for manual workers $h=m$ at the social optimum. \bigskip

\noindent \textbf{Proposition 1.} 
Suppose Assumption 4 holds. Then, at the constrained efficient allocation, for any period \(t\geq 1\),
\[\frac{\partial \tilde{F}_t^*}{\partial K_t^*}>\frac{\partial\tilde{F}_t^*}{\partial AI_t^*},
\]
where \(\tilde{F}_t^*=\tilde{F}(L_{c,t}^*,L_{m,t}^*,K_t^*,AI_t^*)\). \bigskip

\textbf{Proof.} Let \(\lambda_t\) denote the multiplier on the period \(t\) feasibility constraint and \(\mu\) denote the multiplier on the cognitive agents' ICC. Then, the first-order conditions with respect to \(K_t\) and \(AI_t\) are
\begin{equation}
    K_t\colon\,\,\lambda_{t-1}^*=\beta\lambda_t^*\frac{\partial \tilde{F}_t^*}{\partial K_t^*}+\beta X_t^K,\label{4}
\end{equation}
\begin{equation}
    AI_t\colon\,\,\lambda_{t-1}^*=\beta\lambda_t^*\frac{\partial\tilde{F}_t^*}{\partial AI_t^*}+\beta X_t^{AI},\label{5}
\end{equation}
where
\[X_t^K=\mu^*\nu'\bigg(\frac{l_{m,t}^*w_{m,t}^*}{w_{c,t}^*}\bigg)l_{m,t}^*\frac{\partial \big(\frac{w_{m,t}^*}{w_{c,t}^*}\big)}{\partial K_t^*},\]
\[X_t^{AI}=\mu^*\nu'\bigg(\frac{l_{m,t}^*w_{m,t}^*}{w_{c,t}^*}\bigg)l_{m,t}^*\frac{\partial \big(\frac{w_{m,t}^*}{w_{c,t}^*}\big)}{\partial AI_t^*}.\]
By Assumption 1, \({\partial \big({w_{m,t}^*}/{w_{c,t}^*}\big)/}{\partial K_t^*}<0\) and, since \(\mu^*>0\) (due to the binding ICC) and \(\nu'(\cdot)>0\), \(X_t^K<0\). By Assumption 2, \({\partial \big({w_{m,t}^*}/{w_{c,t}^*}\big)/}{\partial AI_t^*}>0\) and \(X_t^{AI}>0\). This, together with the first-order conditions, establishes the result. \hfill $\square$ \bigskip

Traditional capital \(K\) assumes the role of equipment capital from \cite{Slavik2014}: its increase decreases the wage ratio \(w_{m,t}^*/w_{c,t}^*\), making it more attractive for cognitive workers to mimic manual ones. 
However, in our case, an increase of \(AI\) capital raises \(w_{m,t}^*/w_{c,t}^*\), which makes such mimicking less attractive. Thus, at the optimum, the marginal return on \(K\) is higher than the marginal return on \(AI\), i.e., traditional capital should be taxed at a higher rate than AI capital. 

Further following \cite{Slavik2014}, we define the intertemporal wedge that the social planner creates for an agent of type \(h\) for the capital of type \(i\in\{K,AI\}\) from period \(t\) to \(t+1\) as
\[\tau_{i,t+1}(h)=1-\frac{u'(c_{h,t})}{\beta u'(c_{h,t+1})\frac{\partial \tilde{F}_{t+1}^*}{\partial i}}.\]
The intratemporal wedge for agent \(h\) in period \(t\) is
\[\tau_{y,t}(h)=1-\frac{\nu'(l_{h,t})}{w_{h,t}u'(c_{h,t})}.\] \bigskip

\noindent \textbf{Proposition 2.} 
Suppose Assumption 4 holds. Then, in all periods $t\geq1$, the optimal wedge on traditional capital $K$ is strictly positive and independent of agent type, whereas the optimal wedge on $AI$ capital is strictly negative and also independent of agent type. Thus, for every \(h\in H\),
\[ \tau_{K,t}^\ast \equiv \tau_{K,t}^\ast(h) > 0 > \tau_{AI,t}^\ast \equiv \tau_{AI,t}^\ast(h)
.\] \bigskip

\textbf{Proof.} The first-order optimality conditions with respect to \(c_{c,t}\) and \(c_{m,t}\) are
\[c_{c,t}\colon\,\,u'(c_{c,t})\big[\pi_c+\mu^*\big]={\lambda_t^*\pi_c},\]
\[c_{m,t}\colon\,\,u'(c_{m,t})\big[\pi_m-\mu^*\big]={\lambda_t^* \pi_m}.\]
Thus, for any \(h\in H\) 
\[\frac{\lambda_{t-1}^*}{\lambda_{t}^*}=\frac{u'(c_{h,t-1})}{u'(c_{h,t})}.\]
Combining this with conditions (4) and (5), we get
\[\tau_{i,t}^*(h)=\frac{-X_{t}^i}{\lambda_{t}^*\frac{\partial F_t^*}{\partial i}}.\]
The final result follows from \(\lambda_t^*\big(\partial F_t^*/\partial i)>0\), \(X_t^K<0\), and \(X_t^{AI}>0\). \hfill $\square$ \bigskip

According to our result, optimality requires a tax on traditional capital as in \cite{Slavik2014} and a subsidy for $AI$, which is a novel insight. \bigskip

\noindent \textbf{Proposition 3.} Suppose Assumption 4 holds. Then, the optimal intratemporal wedge of cognitive agents is negative, \(\tau_{y,t}^*(c)<0\). \bigskip

\textbf{Proof.} The first-order conditions with respect to \(c_{c,t}\) and \(l_{c,t}\) are
\[c_{c,t}\colon\,\,u'(c_{c,t}^*)\big[\pi_c+\mu^*\big]={\lambda_t^*\pi_c},\]
\[l_{c,t}\colon\,\,\nu'(l_{c,t}^*)\big[\pi_c+\mu^*-\frac{Y_t^*}{\nu'(l_{c,t}^*)}\big]={\lambda_t^*\pi_cw_{c,t}^*},\]
where 
\[Y_t=\mu^*\nu'\bigg(\frac{l_{m,t}^*w_{m,t}^*}{w_{c,t}^*}\bigg)l_{m,t}^*\frac{\partial\big(\frac{w_{m,t}^*}{w_{c,t}^*}\big)}{\partial L_{c,m}^*}z_c\pi_c.\]
By Assumption 3, \(Y_t>0\). From the first-order conditions, \(\pi_c+\mu^*-\frac{Y_t^*}{\nu'(l_{c,t}^*)}>0\). Combining the first-order conditions yields
\[\tau_{y,t}^*(c)=1-\frac{\pi_c+\mu^*}{\pi_c+\mu^*-\frac{Y_t}{\nu'(l_{c,t}^*)}}<0,\]
which establishes the proof. \hfill $\square$ \bigskip

This result implies that subsidizing cognitive work is optimal, as in \cite{Slavik2014}. Increasing the supply of cognitive work decreases the cognitive wage premium and raises $w_m/w_c$. This makes mimicking less profitable for cognitive-type agents. While it seems that the intratemporal wedge of manual agents should be positive, this is not necessarily true if \(w_m>w_c\).  

\subsection{ICC Binding for Manual Workers}

\noindent \textbf{Assumption $4^\prime$:} 
 The ICC (\ref{eq:3}) binds for manual workers $h=m$ and is slack for cognitive workers $h=c$ at the social optimum. \bigskip
 
\noindent\textbf{Proposition 1'.} Suppose Assumption 4' holds. Then, at the constrained efficient allocation, for any period \(t\geq 1\),
\[\frac{\partial \tilde{F}_t^*}{\partial AI_t^*}>\frac{\partial \tilde{F}_t^*}{\partial K_t^*}.\] \bigskip

\textbf{Proof.} Analogous to the proof of Proposition 1. \hfill $\square$ \bigskip

\noindent\textbf{Proposition 2'.} Suppose Assumption 4' holds. Then, in all periods $t\geq1$, the optimal wedge on AI capital is strictly positive and independent of agent type, whereas the optimal wedge on traditional capital is strictly negative and also independent of agent type. That is, for every \(h\in H\),
\[ \tau_{AI,t}^\ast \equiv \tau_{AI,t}^\ast(h) > 0 > \tau_{K,t}^\ast \equiv \tau_{K,t}^\ast(h)
.\] \bigskip

\textbf{Proof.} Analogous to the proof of Proposition 2. \hfill $\square$ \bigskip

\noindent\textbf{Proposition 3'.} Suppose Assumption 4' holds. Then the optimal intratemporal wedge of manual agents is negative, \(\tau_{y,t}^*(m)<0\). \bigskip

\textbf{Proof.} Analogous to the proof of Proposition 3. \hfill $\square$ \bigskip

The change in Assumption 4 reverses the results of all propositions. When cognitive agents are those who find mimicking attractive (their ICC is slack), taxing AI and subsidizing traditional capital is optimal from the social planner perspective, as is subsidizing manual work.


One policy recommendation that is often suggested for the age of AI is the introduction of universal basic income (UBI). If UBI is introduced that redistributes the tax revenues in a lump-sum manner, our results do not change qualitatively. To see this, consider the provision of a baseline consumption level of \(UBI_t\) for everybody. Then total consumption of agent \(h\) in period \(t\) is given by \(\tilde{c}_{h,t}=\bar{c}_{h,t}+UBI_t\), where \(\bar{c}_{h,t}\) is chosen by the social planner. This change has no impact on Propositions 1 and 1', because UBI neither alters the production technology nor the ICCs. The qualitative results of Propositions 2, 2', 3 and 3' also remain the same. We can summarize this in the following remark. 

\begin{remark}

The introduction of UBI 
does not change our qualitative findings as reflected in Propositions 1-3 and 1'-3'.

\end{remark}



\section{Conclusion}

When AI becomes a good substitute for human cognitive work and its adoption drives down the wages of cognitive workers relative to manual workers, cognitive workers may find it optimal to seek manual jobs. In such a scenario, it is optimal to tax AI capital and to subsidize traditional capital and manual labor. This is an important new result of dynamically optimal taxation for the age of AI. 
%
The potential reversal of manual vs. cognitive workers' incentives constitutes another critical threshold in AI development, other than human-level AI performance at benchmarks or the threshold where human cognitive work and AI become gross substitutes.

\newpage

\bibliographystyle{apalike}
\bibliography{References_7}

\end{document}